\documentclass{ieeeaccess}
\usepackage{cite}
\usepackage{amsmath,amssymb,amsfonts}
\usepackage{algorithmic}
\usepackage{graphicx}
\usepackage{textcomp}

\usepackage{multirow}
\usepackage{url}
\usepackage{braket}
\usepackage{booktabs}
\usepackage{threeparttable}
\usepackage{color}

\usepackage{caption}
\usepackage{framed}

\def\BibTeX{{\textrm B\kern-.05em{\sc i\kern-.025em b}\kern-.08em
    T\kern-.1667em\lower.7ex\hbox{E}\kern-.125emX}}

\begin{document}
\history{Date of publication xxxx 00, 0000, date of current version xxxx 00, 0000.}
\doi{10.1109/TQE.2020.DOI}

\onecolumn
\begin{framed}
   \noindent
   This work has been submitted to the IEEE for possible publication. Copyright may be transferred without notice, after which this version may no longer be accessible.%
\end{framed}
\clearpage
\twocolumn

\title{Effectiveness of Hybrid Optimization Method for Quantum Annealing Machines}

\author{
\uppercase{Shuta Kikuchi}\authorrefmark{1,2},
\uppercase{Nozomu Togawa}\authorrefmark{3}, \IEEEmembership{Member, IEEE},
\uppercase{and Shu Tanaka}\authorrefmark{1,2,4,5}, \IEEEmembership{Member, IEEE}
}

\address[1]{Graduate School of Science and Technology, Keio University, Yokohama, Kanagawa 223-8522, Japan}
\address[2]{Keio University Sustainable Quantum Artificial Intelligence Center (KSQAIC), Keio University, Minato-ku, Tokyo 108-8345, Japan}
\address[3]{Department of Computer Science and Communications Engineering, Waseda University, Shinjuku-ku, Tokyo 169-8555, Japan}
\address[4]{Department of Applied Physics and Physico-Informatics, Keio University, Yokohama, Kanagawa 223-8522, Japan}
\address[5]{Human Biology-Microbiome-Quantum Research Center (WPI-Bio2Q), Keio University, Minato-ku, Tokyo 108-8345, Japan}

\tfootnote{
This work was partially supported by the Japan Society for the Promotion of Science (JSPS) KAKENHI (Grant Number JP23H05447), the Council for Science, Technology, and Innovation (CSTI) through the Cross-ministerial Strategic Innovation Promotion Program (SIP), ``Promoting the application of advanced quantum technology platforms to social issues'' (Funding agency: QST), Japan Science and Technology Agency (JST) (Grant Number JPMJPF2221).}

\markboth
{Kikuchi \headeretal: Effectiveness of Hybrid Optimization Method for Quantum Annealing Machines}
{Kikuchi \headeretal: Effectiveness of Hybrid Optimization Method for Quantum Annealing Machines}

\corresp{Corresponding author: Shuta Kikuchi (e-mail: kikuchi.shuta@keio.jp).}

\begin{abstract}
To enhance the performance of quantum annealing machines, several methods have been proposed to reduce the number of spins by fixing spin values through preprocessing.
We proposed a hybrid optimization method that combines a simulated annealing (SA)-based non-quantum-type Ising machine with a quantum annealing machine. 
However, its applicability remains unclear. 
Therefore, we evaluated the performance of the hybrid method on large-size Ising models and analyzed its characteristics. 
The results indicate that the hybrid method improves upon solutions obtained by the preprocessing SA, even if the Ising models cannot be embedded in the quantum annealing machine.
We analyzed the method from three perspectives: preprocessing, spin-fixed sub-Ising model generation method, and the accuracy of the quantum annealing machine.
From the viewpoint of the minimum energy gap, we found that solving the sub-Ising model with a quantum annealing machine results in a higher solution accuracy than solving the original Ising model. 
Additionally, we demonstrated that the number of fixed spins and the accuracy of the quantum annealing machine affect the dependency of the solution accuracy on the sub-Ising model size.
\end{abstract}

\begin{keywords}
Ising machine, Ising model, quantum annealing, simulated annealing, fixing spin, spin reduction.
\end{keywords}

\titlepgskip=-15pt

\maketitle
\section{Introduction}
\label{sec:introduction}
\PARstart{C}{ombinatorial} optimization problems are a category of problems that seek the optimal combination of decision variables to minimize or maximize an objective function while satisfying a set of constraints. 
Many such problems are known as nondeterministic polynomial time (NP)-complete or NP-hard problems in computational complexity theory~\cite{Karp1972reducibility}.
The number of candidate solutions to these problems increases exponentially, with the number of decision variables.
Because these problems appear in various social and industrial situations, there is growing interest in developing technologies to obtain optimal or quasi-optimal solutions efficiently and accurately.

Ising machines have attracted considerable attention as accurate and efficient solvers for combinatorial optimization problems~\cite{mohseni2022ising}.
Many research studies applying Ising machines have been reported across numerous applications domains~\cite{yarkoni2022quantum}, including traffic optimization~\cite{neukart2017traffic, irie2019quantum, ohzeki2019control, Bao2021-a, Bao2021-b, Mukasa2021, haba2022travel, noguchi2025hybrid}, quantum compilation~\cite{naito2023isaaq}, material simulations~\cite{king2018observation, harris2018phase, utimula2021quantum, endo2022phase, sampei2023quantum, honda2024development, xu2025quantum, aoki2025formulation}, portfolio optimization~\cite{rosenberg2016solving, Tanahashi2019, tatsumura2023real}, biology~\cite{degasperi2017performance, li2018quantum}, and black-box optimization~\cite{kitai2020designing, izawa2022continuous, inoue2022towards, seki2022black, tucs2023quantum, nawa2023quantum}.

\begin{table*}[t]
\centering
\begin{threeparttable}
\caption{Overview of Some Implemented Ising Machines}
\label{table:Ising_machine_spec}
\begin{tabular} {cccccc} \toprule
Ising machine & Spin size & Topology & Signed bit-width & Internal algorithm & Hardware \\ \midrule
CMOS annealing machine~\cite{yamaoka201620k} & 20,480 & 3-D lattice & 2 bits & SA & ASIC \\
CMOS annealing machine~\cite{takemoto20192} & 61,952 & Kings & 3 bits & SA & ASIC \\
CMOS annealing machine~\cite{okuyama2017ising} & 2,304 & Kings & 8 bits & SQA & FPGA \\
CMOS annealing machine~\cite{okuyama2019binary} & 100,000 & Complete & 10 bits & MA & GPU \\
Digital Annealer~\cite{tsukamoto2017accelerator, matsubara2018ising, aramon2019physics} & 1,024 & Complete & 26 bits & SA/PT & FPGA \\
Digital Annealer~\cite{aramon2019physics, matsubara2020digital} & 8,192 & Complete & 64 bits & SA/PT & ASIC \\
Simulated Bifurcation (SB) machine~\cite{tatsumura2019fpga} & 4,096 & Complete & 32 bits$^*$ & SB & FPGA \\
Simulated Bifurcation (SB) machine~\cite{goto2021high} & 100,000 & Complete & 32 bits$^*$ & bSB/dSB & GPU \\
Fixstars Amplify AE~\cite{FixAE} & 262,144 & Complete & 32/64 bits & SA & GPU \\
Coherent Ising machine~\cite{honjo2021100} & 100,000 & Complete & 2 bits & $-$ & Laser, FPGA \\
D-Wave 2000Q~\cite{king2018observation} & 2,048 & Chimera & $-$ & QA & SQUID \\
D-Wave Advantage~\cite{boothby2020next, mcgeoch2021advantage} & 5,627 & Pegasus & $-$ & QA & SQUID \\ \bottomrule
\end{tabular}
\begin{tablenotes}\footnotesize
\item[]SQA: Simulated Quantum Annealing, MA: Momentum Annealing, PT: Parallel Tempering, bSB: ballistic SB, dSB: discrete SB
\item[] $^*$ Single precision floating point
\end{tablenotes}
\end{threeparttable}
\end{table*}

Table~\ref{table:Ising_machine_spec} summarizes the implemented Ising machines.
Ising machines can be categorized into two types.
The first type is quantum annealing machines such as D-Wave Advantage~\cite{mcgeoch2021advantage}.
These are implemented using superconducting quantum interference devices (SQUIDs) and their internal algorithm is quantum annealing (QA)~\cite{kadowaki1998quantum, farhi2000quantum, farhi2001quantum}.
Quantum annealing machines are expected to have an advantage over classical optimizers and may provide quantum speedup for certain classes of problems~\cite{kim2025quantum}.
The second type is non-quantum-type Ising machines, which are also referred to as a simulated-annealing-based Ising machine, such as complementary metal-oxide-semiconductor (CMOS) annealing machines~\cite{yamaoka201620k, takemoto20192, okuyama2017ising, okuyama2019binary}, Digital Annealer~\cite{tsukamoto2017accelerator, matsubara2018ising, matsubara2020digital}, and the Fixstars Amplify Annealing Engine (AE)~\cite{FixAE}.
They are implemented using digital circuits such as application-specific integrated circuits (ASICs), field-programmable gate arrays (FPGAs), and graphics processing units (GPUs).
The basis of their internal algorithm is simulated annealing (SA)~\cite{kirkpatrick1983optimization,johnson1989optimization,johnson1991optimization, isakov2015optimised}.
The current advantage of non-quantum-type Ising machines is that they can handle larger spin sizes than quantum annealing machines.

Before solving combinatorial optimization problems with an Ising machine, two operations are required.
First, combinatorial optimization problems are mapped to the \textit{Ising model}~\cite{Ising} or its equivalent model called the \textit{quadratic unconstrained binary optimization (QUBO) model}.
QUBO is also referred to as an \textit{unconstrained binary quadratic programming (UBQP) problem}~\cite{kochenberger2014unconstrained}.
Most problems can be mapped to the Ising model or QUBO~\cite{lucas2014ising, Tanaka2017}.
Ising machines search for low-energy states of the Ising model or QUBO.
An Ising model is defined on an undirected graph $G=(V, E)$, where $V$ and $E$ are the sets of vertices and edges, respectively.
The Hamiltonian $\mathcal{H}$ of the Ising model is defined as
\begin{align}
\mathcal{H} = - \sum_{i\in V}h_{i}\sigma_{i} - \sum_{(i,j)\in E} J_{i, j}\sigma_{i}\sigma_{j} ,
\label{eq:H}
\end{align}
where $\sigma_i \in \{+1, -1\}$ denotes the spin on vertex $i \in V$.
The coefficients of the RHS, $h_{i}$ and $J_{i,j}$, are the magnetic field on vertex $i \in V$ and the interaction on edge $(i, j) \in E$, respectively.
The formulated Ising model is referred to as the logical Ising model.

Second, the logical Ising model is mapped onto a physical Ising model that corresponds to the hardware limitations of the Ising machine.
Hardware limitations for an Ising machine include topology limitations, which refer to the network structure of interactions between spins; bit-width limitations, which refer to the range of input coefficients; and spin-size limitations, which refer to the number of spins that can be input into the machine.
The hardware specifications are presented in Table~\ref{table:Ising_machine_spec}.
Methods to mitigate topology limitations~\cite{choi2011minor, cai2014practical, oku2019fully, okada2019improving, shirai2020guiding} and bit-width limitations~\cite{Oku2020, yachi2023efficient, kikuchi2023dynamical} have been proposed. 
For spin-size limitations, methods have been proposed to effectively reduce the number of variables through appropriate preprocessing and solve small-scale problems using an Ising machine~\cite{qbsolv, karimi2017boosting ,karimi2017effective, irie2021hybrid, atobe2021hybrid, noguchi2023trip, kikuchi2023hybrid, raymond2023hybrid, takabayashi2024hybrid, lee2024statistical, kanai2024annealing}. 

Among the methods for reducing the number of variables, there is an approach called ``sample persistence''~\cite{karimi2017boosting, karimi2017effective}.
This method involves obtaining multiple solutions of the original Ising model before reducing the number of variables.
In sample persistence, variables that consistently hold the same value across these solutions are considered to be stable spins in a low-energy state, whereas spins with varying values are regarded as unstable spins that are difficult to determine.
Several algorithms inspired by this method have been proposed and reported to produce improved performance~\cite{karimi2017boosting, karimi2017effective, atobe2021hybrid, noguchi2023trip, kikuchi2023hybrid, lee2024statistical}.
However, to the best of our knowledge, no analysis has been conducted on the phenomena that affect the solution accuracy achieved by this method.

As one of these algorithms, a previous study proposed a hybrid optimization method that combines the advantages of a quantum annealing machine and a non-quantum-type Ising machine~\cite{kikuchi2023hybrid}.
In this hybrid optimization method, multiple solutions are obtained using a non-quantum-type Ising machine first. 
Then, based on the concept of sample persistence, the spins of the Ising model are reduced, and small-scale problems are solved using a quantum annealing machine.
The performance of the hybrid method for fully connected Ising models that can be input into the D-Wave Advantage was evaluated using simulations with SA and the D-Wave Advantage.
The hybrid optimization method outperformed D-Wave Advantage and the preprocessing SA alone.
This method can potentially mitigate the spin-size limitation of a quantum annealing machine because the size of the sub-Ising model input into the quantum annealing machine is smaller than that of the original Ising model. 
However, the performance of the method for Ising models that cannot be input into D-Wave Advantage has not been evaluated.

Therefore, in this study, we investigated the performance of a hybrid optimization method for large-scale Ising models that cannot be input into a quantum annealing machine.
The effects of the hybrid method parameters were also investigated.
Moreover, we analyzed the characteristics of the hybrid method from the perspective of preprocessing, sub-Ising model generation method, and accuracy of the quantum annealing machine.

The rest of this paper is organized as follows. 
Section~\ref{sec:hybrid_method} introduces the hybrid optimization method.
Section~\ref{sec:internal_algorithms} introduces the internal algorithms of Ising machines.
Section~\ref{sec:evaluation_HM} presents simulations using SA as a non-quantum-type Ising machine and D-Wave Advantage as a quantum annealing machine, with a fully connected Ising model that cannot be input into D-Wave Advantage.
Section~\ref{sec:analysis_HM} presents an analysis of the hybrid optimization method using simulations of SA and QA implemented using the master equation and the Schrödinger equation.
Section~\ref{sec:conclusion} presents the conclusions of the study. 
The appendices provide supplemental information on the relationship between minimum energy gaps of sub-Ising model with unstable spins and sub-Ising models size (Appendix~\ref{sec:appendixA}), analysis of SA and minimum energy gaps (Appendix~\ref{sec:appendixB}), and evaluation of the effect of annealing time (Appendix~\ref{sec:appendixC}).
\section{Hybrid optimization method}
\label{sec:hybrid_method}
A hybrid optimization method combining a non-quantum-type-Ising machine and a quantum annealing machine was proposed~\cite{kikuchi2023hybrid}.
In this section, we describe the method for generating the sub-Ising model in Subsection~\ref{subsec:sub-ising} and the algorithm of the hybrid optimization method in Subsection~\ref{subsec:algorithm}.
\subsection{Sub-Ising model generation}
\label{subsec:sub-ising}
Sub-Ising models are generated by fixing a subset of spins from the original Ising model to either $+1$ or $-1$.
The values of fixed spins are determined using tentative solutions. 

Here, let $n$ and $m$ be the number of spins in the original Ising model and sub-Ising model, respectively.
The spins of the original Ising model are denoted as $S = \{ \sigma_1$, $\sigma_2$, $\cdots$, $\sigma_n \}$.
In addition, $\tilde{\sigma}_i =$ ($\tilde{\sigma}_1$, $\tilde{\sigma}_2$, $\cdots$, $\tilde{\sigma}_n$) denotes the spin state of a tentative solution, where $\tilde{\sigma}_i$ is either $-1$ or $1$.
Consequently, if $S'$ denotes a subset of the spins of the sub-Ising model, the Hamiltonian of the sub-Ising model $\mathcal{H}_\mathrm{sub}$ is expressed as follows:
\begin{align}
  \mathcal{H}_\mathrm{sub} = - \sum_{\substack{i \\ \sigma_{i}\in S'}}R_{i}\sigma_{i} - \sum_{\substack{(i,j) \\ \sigma_{i},\sigma_{j}\in S'}} J_{i,j}\sigma_{i}\sigma_{j} + C ,
  \label{eq:H_sub}
\end{align}
where
\begin{align}
  R_{i} = h_{i} + \sum_{\substack{j \\ \tilde{\sigma}_{j}\notin S'}} J_{i,j}\tilde{\sigma}_{j},
  \label{eq:L_i}
\end{align}
\begin{align}
  C = - \sum_{\substack{i \\ \tilde{\sigma}_{i} \notin S'}} h_{i} \tilde{\sigma}_{i} - \sum_{\substack{(i,j) \\ \tilde{\sigma}_{i},\tilde{\sigma}_{j}\notin S'}} J_{i,j}\tilde{\sigma}_{i}\tilde{\sigma}_{j}.
  \label{eq:C}
\end{align}
Here, $C$ comprises the contributions from spins fixed by the tentative solution values and their associated coefficients. 
We refer to $C$ as the sub-Ising model constant.
\subsection{Algorithm}
\label{subsec:algorithm}
\begin{figure}[t]
    \centering
    \includegraphics[clip,width=0.9\linewidth]{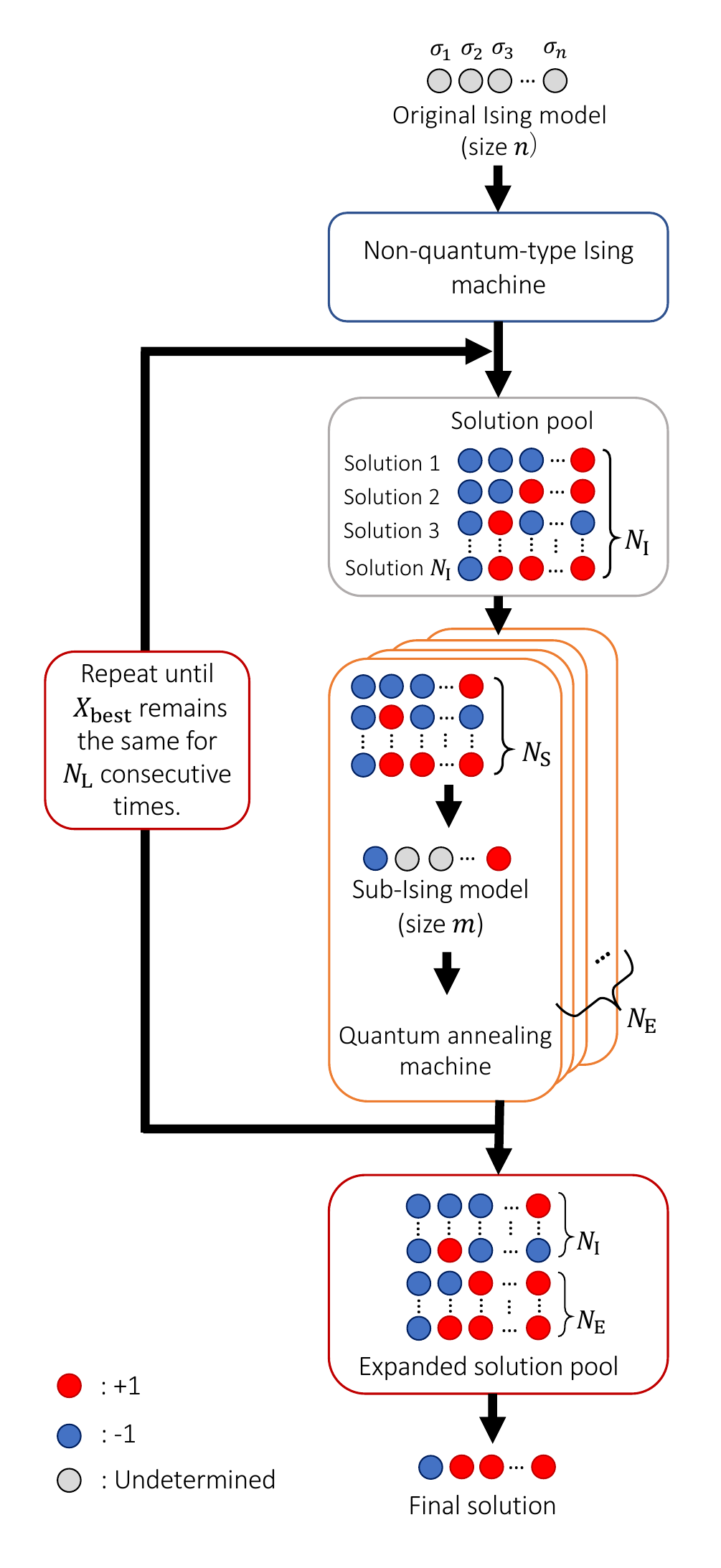}
    \caption{Flowchart of the hybrid optimization method. The red, blue, and gray circles indicate that the spin has $+1$, $-1$, and undetermined values, respectively. The black arrows denote the flow.}
    \label{fig:hybrid_method_scheme}
\end{figure}

In this subsection, we describe the algorithm of the hybrid optimization method in detail. 
Fig.~\ref{fig:hybrid_method_scheme} shows a flowchart of the method, which consists of the following five steps.

\begin{figure}[t]
  \centering
  \includegraphics[clip,width=0.8\linewidth]{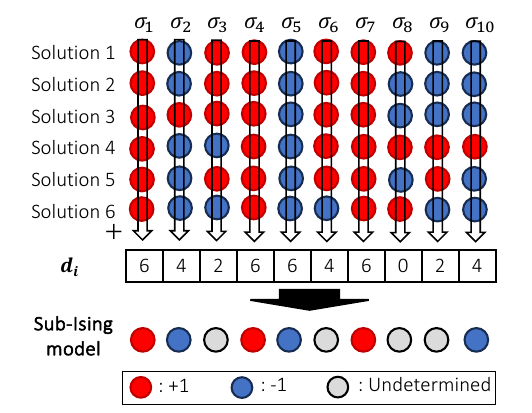}
  \caption{Example of sub-Ising model generation. The red, blue, and gray circles indicate that the spin has $+1$, $-1$, and undetermined values.}
  \label{fig:sub-ising_generation}
\end{figure}

\begin{description}
\item[Step 1:]$N_{\textrm I}$ solutions are obtained using a non-quantum-type Ising machine. These solutions form a solution pool. These solutions are expected to have different spin states because a non-quantum-type Ising machine exhibits stochastic behavior. The solution pool may contain duplicate solutions. The solution with the lowest energy within the solution pool is temporarily selected as the lowest-energy solution $X_{\textrm{best}}$. 
\item[Step 2:]$N_{\textrm S}$ solutions are randomly selected from the solution pool, allowing for duplicates. A sub-Ising model is generated from $N_{\textrm S}$ solutions using the following method (Fig.~\ref{fig:sub-ising_generation}). To select the fixed spins, $d_i$ is calculated for all spins. Here, $d_i$ is given by the following equation:
\begin{align}
d_{i}=\left|\sum^{N_{\textrm S}}_{k=1}\sigma_{k, i}\right|,
\label{eq:d_i}
\end{align}
where $\sigma_{k, i}$ represents the $i$-th spin in the $k$-th solution. When $N_{\textrm S}$ is even or odd respectively, the minimum value of $d_i$ is $0$ or $1$. When a spin indicates the minimum value of $d_i$, it becomes the most unstable. Meanwhile, when $d_i$ is $N_{\textrm S}$, the spin is considered stable. To obtain more unstable spins, spins are collected in the ascending order of $d_i$. The number of collected spins is set to a sub-Ising model size $m$. A tentative solution is randomly chosen from the $N_{\textrm S}$ solutions. Using these components, we generate the sub-Ising model from Eqs.~\eqref{eq:H_sub}--\eqref{eq:C}.
Subsequently, a quantum annealing machine searches for lower-energy states of the sub-Ising model.
\item[Step 3:]Step $2$ is repeated for $N_{\textrm E}$ times. New $N_{\textrm E}$ solutions are added to the solution pool.
\item[Step 4:]From the expanded solution pool, $N_{\textrm I}$ solutions are collected in ascending order of energy, and generate a new solution pool. Within the new solution pool, the new lowest-energy state $X_{\textrm{best}}$ is updated.
\item[Step 5:]The flow from Step $2$--$4$ is repeated until the solutions in the solution pool converge. The repetition continues until $X_{\textrm{best}}$ remains unchanged for $N_{\textrm L}$ consecutive iterations. The final $X_{\textrm{best}}$ is the solution obtained using the hybrid optimization method.
\end{description}

\section{Internal algorithms of Ising machines}
\label{sec:internal_algorithms}
SA and QA are the major underlying internal algorithms implemented in Ising machines.
In this section, we describe these algorithms for implementation as simulations.
\subsection{Simulated Annealing}
\label{subsec:SA}
SA is a metaheuristic that involves introducing the concept of temperature to induce state transitions in the Ising model~\cite{kirkpatrick1983optimization,johnson1989optimization,johnson1991optimization, isakov2015optimised}. 
In SA, the ground state is explored by decreasing the temperature sufficiently slowly from high to low.

SA has been implemented using single-spin-flip Markov Chain Monte Carlo (MCMC). 
For an $n$-spin Ising model, the algorithm is performed as follows:

\begin{description}
\item[Step 1:]A random initial spin state is prepared. 
\item[Step 2:]The initial temperature is set. The temperature is sufficiently high for the Hamiltonian of the Ising model. 
\item[Step 3:]One spin is selected randomly from the $n$ spins.
\item[Step 4:]The chosen spin is flipped according to the transition probability $W(\Delta{\mathcal{H}}, T)$, where $\Delta \mathcal{H}$ and $T$ are the energy difference and temperature, respectively. The energy difference $\Delta \mathcal{H}$ is defined by $\Delta{\mathcal{H}}=\mathcal{H}_\textrm{candidate}-\mathcal{H}_\textrm{current}$, where $\mathcal{H}_\textrm{candidate}$ is the energy of the candidate state in which the chosen spin is flipped, and $\mathcal{H}_\textrm{current}$ is the energy of the current state. 
\item[Step 5:]Steps $3$--$4$ are repeated ``inner loop'' times. The inner loop is typically set to the number of spins $n$. 
\item[Step 6:]The temperature $T$ is decreased and the algorithm returns to Step $3$. 
\item[Step 7:]Step $6$ is repeated ``outer loop'' times. 
\end{description}

Two methods are known for the transition probability: the heat-bath method~\cite{metropolis1953equation, hastings1970monte} and the Metropolis method~\cite{glauber1963time}.
In this study, we used the heat-bath method given by $ \left\{ 1 + \exp \left[ {\Delta \mathcal{H}} / {T(t)} \right] \right\}^{-1}$, where $T(t)$ represents the temperature at time $t$.
The Geman--Geman theorem ensures that SA obtains a ground state when the temperature decreases sufficiently slowly~\cite{Geman1984}.

SA is based on stochastic dynamics called the master equation.
The master equation is described as follows:
\begin{align}
    \frac{dP(\{ \sigma_{i} \}, t)}{dt} = \sum_{j}\mathcal{L}_{i,j}P(\{ \sigma_{j} \}, t),
    \label{eq:SA_master_equation}
\end{align}
where $P(\{ \sigma_{i} \}, t)$ represents the probability of the spin state $\{ \sigma_{i} \}$ at time $t$. 
We consider the conditions for the transitions from the spin state $\{\sigma_j\}$ to $\{\sigma_i\}$ in \eqref{eq:SA_master_equation}.
Using this differential equation as the master equation allows us to obtain the probability of all spin states when SA is performed.
To perform single-spin flips, the elements of transition matrix $\mathcal{L}_{i,j}$ are given as
\begin{align}
    \mathcal{L}_{i,j} =
    \begin{cases}
    \left\{ 1 + \exp \left[ \dfrac{\Delta \mathcal{H}}{T(t)} \right] \right\}^{-1} & \text{(single-spin difference)}\\
    \displaystyle -\sum_{l \neq i} \mathcal{L}_{l,i} & \text{(} i = j \text{)}\\
    0 & \text{(otherwise)}
    \end{cases}
    .
    \label{eq:SA_master_equation_case}
\end{align}
\subsection{Quantum Annealing}
\label{subsec:QA}
QA is a metaheuristic that uses quantum effects instead of temperature fluctuations, as in SA~\cite{kadowaki1998quantum, farhi2000quantum, farhi2001quantum}.
QA is performed using the following time-dependent Hamiltonian:
\begin{align}
  \hat{\mathcal{H}}\left(t\right) = \frac{t}{\tau} \hat{\mathcal{H}}_0 + \left( 1 - \frac{t}{\tau} \right) \hat{\mathcal{H}}_{\textrm q},
  \label{eq:H_QA}
\end{align}
\begin{align}
  \hat{\mathcal{H}}_0 = - \sum_{i\in V}h_{i}\hat{\sigma}_{i}^{z} - \sum_{(i,j)\in E} J_{i,j}\hat{\sigma}_{i}^{z}\hat{\sigma}_{j}^{z},
  \label{eq:H_0}
\end{align}
\begin{align}
  \hat{\mathcal{H}}_{\textrm q} = - \sum_{i\in V}\hat{\sigma}_{i}^{x},
  \label{eq:H_q}
\end{align}
where $\hat{\sigma}_{i}^{z}$ and $\hat{\sigma}_{i}^{x}$ are the Pauli matrices of vertex $i$.
Equation~\eqref{eq:H_0} corresponds to the Hamiltonian of the Ising model in Eq.~\eqref{eq:H}, and Eq.~\eqref{eq:H_q} is the Hamiltonian that causes the quantum effect.
Here, $\tau$ is the annealing time, with $0 \leq t \leq \tau$.
The initial state at $t=0$ is $\hat{\mathcal{H}}(0) = \hat{\mathcal{H}}_{\textrm q}$ and the final state at $t=\tau$ is $\hat{\mathcal{H}}(\tau) = \hat{\mathcal{H}}_0$.
In states with significant quantum fluctuations, $\hat{\mathcal{H}}_{\textrm q}$ is approximately represented by a superposition of all states that have almost equal probabilities, which is the ground state.
QA attempts to obtain the nontrivial ground state of $\hat{\mathcal{H}}_0$ by starting from the trivial ground state of $\hat{\mathcal{H}}_{\textrm q}$ and gradually reducing quantum fluctuations.
According to the adiabatic theorem of quantum mechanics~\cite{kato1950adiabatic, messiah1962quantum, jansen2007bounds}, the system remains in its ground state during the evolution if the annealing is performed sufficiently slowly.
The required runtime scales inversely with the square of the minimum energy gap between the ground state and the first excited state.
Since the energy gap is closely related to computational difficulty, the minimum gap is commonly used as an indicator of problem hardness in quantum annealing.
Here, the minimum energy gap refers to the smallest energy gap during the annealing process.

The dynamics of QA are described by the time-dependent Schr\"odinger equation
\begin{align}
  \textrm{i}\frac{d}{dt}\ket{\Psi(t)} = \hat{\mathcal{H}}(t)\ket{\Psi(t)},
  \label{eq:Schrodinger_equation}
\end{align}
where $\ket{\Psi(t)}$ denotes the wave function at time $t$.
Here, we used the natural unit, in which the Dirac constant is set to $\hbar = 1$.
\section{Evaluation of hybrid optimization method}
\label{sec:evaluation_HM}
\subsection{Setup of the experiment}
\label{subsec:setup_HM}
In this evaluation, SA implemented via MCMC was used as a non-quantum-type Ising machine, and D-Wave Advantage was employed as a quantum annealing machine.

The SA parameters were set as described below:
The initial states were set randomly, and the initial temperatures $T_\textrm{initial}$ were set to $\lceil 2v_\mathrm{max} \rceil$ for each instance. 
Here, $v_\textrm{max}$ is the maximum value among $v_{i}$ defined by $v_{i}=\left|h_{i}+\sum_{j \in \partial_{i}}{J_{i,j}}\right|$.
This represents the absolute value of the sum of the local field and interactions for the $i$-th spin, and $\partial_{i}$ denotes the set of adjacent vertices of the vertex $i$.
The initial temperature was set sufficiently high to accommodate the transition.
The temperature schedule was set to power-law decay for every outer loop, which is given by $T(u)=T_\textrm{initial} \times r^u$, where $r$ is the cooling rate and $u$ is the $u$-th outer loop.
The outer loop is set to $50$.
The cooling rate was set such that the final temperature was equal to $0.1$, which was sufficiently low on the energy scale of the original Ising model.
The inner loop is set to the number of spins in the original Ising model.

When we used D-Wave Advantage, a sub-Ising model was annealed $100$ times and the best solution was selected.
The other parameters are set to their default values~\cite{Ocean}.
For example, the annealing time was set to $20$~$\mu$s, and the problem was rescaled such that the magnetic fields were in the range $[-2, 2]$ and the interactions were in the range $[-1, 1]$.

We used a fully connected random Ising model, which is represented as a complete graph.
Because the edge density does not change regardless of which variables are fixed when the sub-Ising model size is the same, we chose a complete graph. 
To avoid trivial degeneracy, the coefficients of the magnetic field and the interaction were randomly selected according to a Gaussian distribution with a mean of zero and a standard deviation of unity.
The sub-Ising model generation method is ineffective when the Ising models exhibit degeneracy. 
Notably, zero was excluded for both the interactions and magnetic fields. 

To evaluate the effectiveness of hybrid optimization methods for large-scale problems, we used Ising models with sizes that could not be embedded in D-Wave Advantage.
D-Wave Advantage has $5,627$ qubits and a Pegasus graph topology~\cite{dattani2019pegasus}.
A complete graph of $177$ spins can be embedded using minor embedding~\cite{DW_update}.
Therefore, we employed the original Ising models with the number of spins set to $240$, $320$, $480$, and $640$.
One instance was prepared for each model with the corresponding number of spins.

For the parameters of the hybrid optimization method, we set $N_{\textrm I} = 20$, $N_{\textrm S} = 10$, $N_{\textrm E} = 20$, and $N_{\textrm L} = 3$.
The sub-Ising model size $m$ was set to $40$, $80$, $120$, and $160$.
\subsection{Numerical experiment}
\label{subsec:numerical_experiment}
\begin{figure*}[ht]
  \centering
  \begin{minipage}{0.4\linewidth}
    \includegraphics[width=\linewidth]{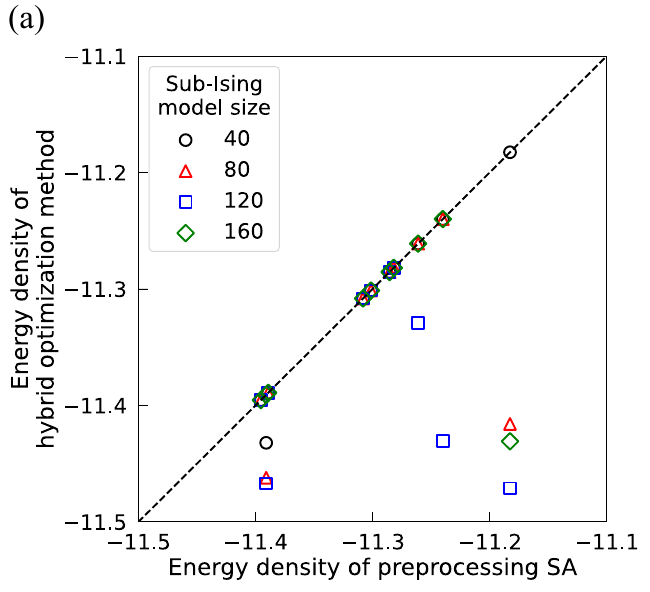}
  \end{minipage}
  \hspace{1cm}
  \begin{minipage}{0.4\linewidth}
    \includegraphics[width=\linewidth]{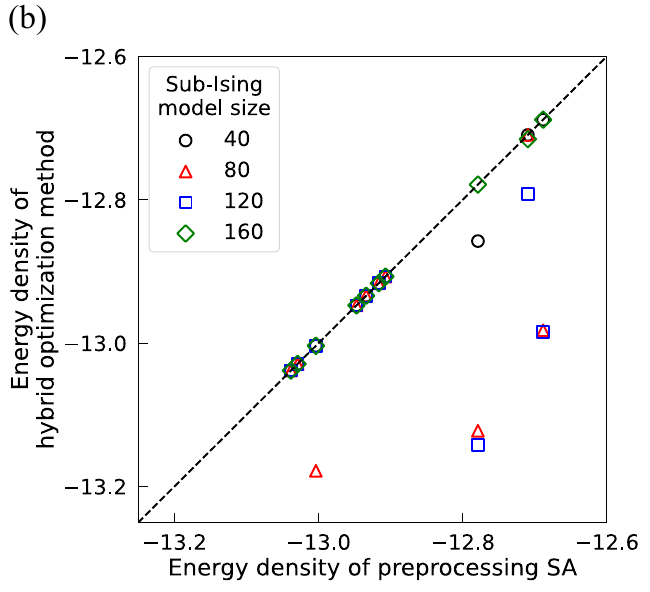}
  \end{minipage}
  \\
  \vspace{5mm}
  \begin{minipage}{0.4\linewidth}
    \includegraphics[width=\linewidth]{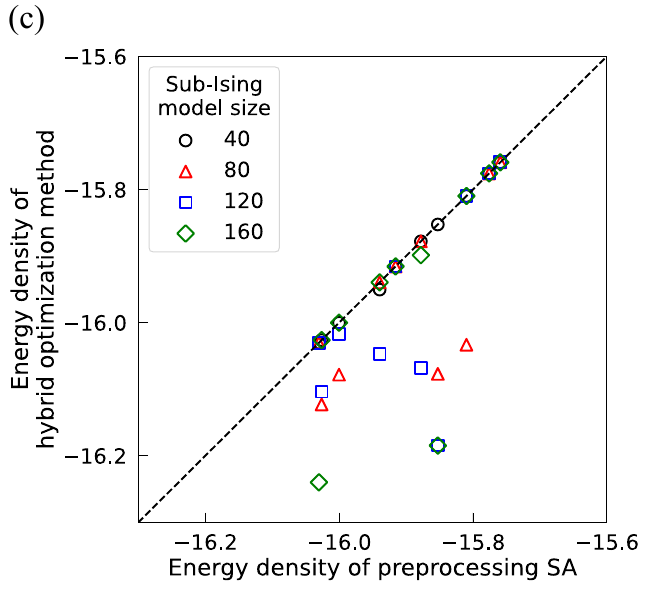}
  \end{minipage}
  \hspace{1cm}
  \begin{minipage}{0.4\linewidth}
    \includegraphics[width=\linewidth]{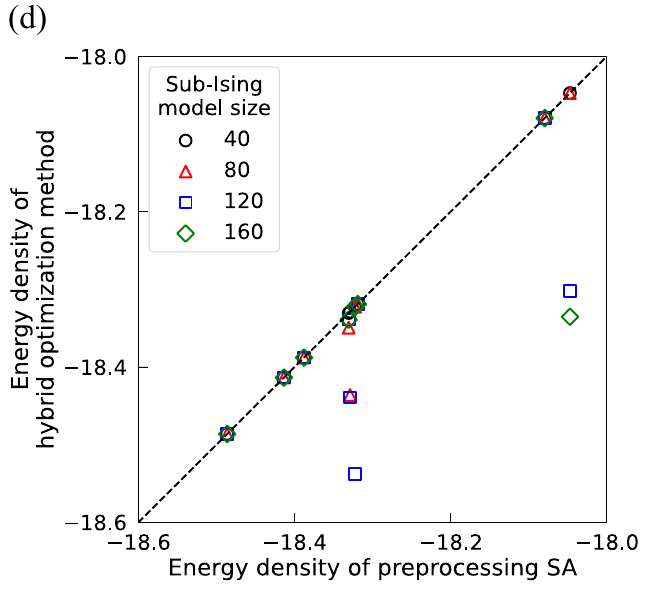}
  \end{minipage}
  \caption{Energy densities of the hybrid optimization method and preprocessing SA. Original Ising model sizes are (a) $n=240$, (b) $n=320$, (c) $n=480$, and (d) $n=640$. The black circles, red triangles, blue squares, and green diamonds denote sub-Ising model sizes $m = 40$, $80$, $120$, and $160$, respectively.}
  \label{fig:Sub_size}
\end{figure*}

We evaluated the updates to solutions obtained from preprocessing SA using the hybrid optimization method for large-scale problems.
Furthermore, the dependency on the sub-Ising model size was also evaluated.
The results are presented in Fig.~\ref{fig:Sub_size}.
The figure shows the energy densities of the solutions obtained by preprocessing SA and the hybrid optimization method.
The energy density is the internal energy per spin (i.e., $\mathcal{H}/n$), and the plotted solutions are $X_{\textrm{best}}$ of the solution pool.
The sets of solution pools obtained by preprocessing SA were the same for all conditions of the sub-Ising model size.
These data were obtained from $10$ simulations.

Fig.~\ref{fig:Sub_size} shows that the hybrid method updates the solutions from those of preprocessing SA, even if the Ising model cannot be embedded in D-Wave Advantage.
Focusing on the sub-Ising model size $m$, it was shown that the solutions are less likely to update when $m = 40$ and $m = 160$.
Similar results were observed in simulations using Ising models with sizes that can be embedded in D-Wave Advantage. 
When the original Ising model size was $160$, the solutions were less likely to be updated when $m$ was small (i.e., $m=14$) or large (i.e., $m=120, 144$), and the solutions were updated when $m = 40$ or $m = 80$~\cite{kikuchi2023hybrid}.
\section{Analysis of hybrid optimization method}
\label{sec:analysis_HM}
In the previous section, we showed the effectiveness of the hybrid optimization method for large-scale original Ising models and the dependency of the sub-Ising model size.
In this section, we analyze the effectiveness and dependency using numerical simulations of SA and QA.

In this study, we use a small-scale Ising model for analysis.
The number of spins was set to eight, and all spins interacted with each other. 
The coefficients were determined by following a Gaussian distribution with a mean of zero and a standard deviation of unity, and the Ising models were prepared in two instances (Ising models A and B).
Upon exploring all spin states in these Ising models, no degeneracy was observed.
\subsection{Fixed spin selection by preprocessing SA}
\label{subsec:equation_SA}
\begin{figure*}[ht]
  \centering
  \begin{minipage}{0.32\linewidth}
    \includegraphics[width=\linewidth]{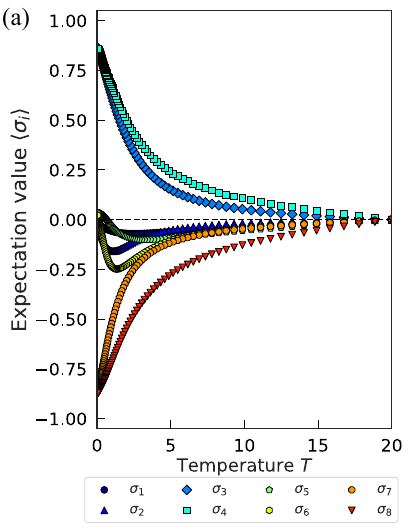}
  \end{minipage}
  \hspace{1cm}
  \begin{minipage}{0.32\linewidth}
    \includegraphics[width=\linewidth]{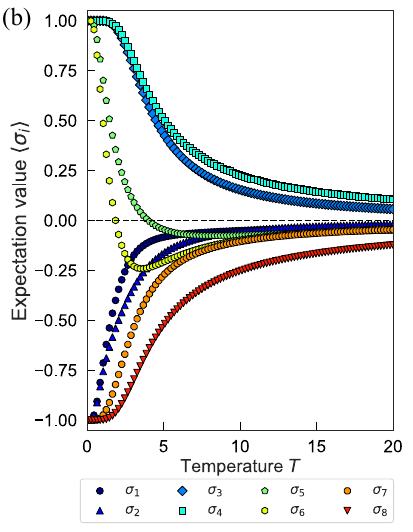}
  \end{minipage}
  \\
  \vspace{5mm}
  \begin{minipage}{0.32\linewidth}
    \includegraphics[width=\linewidth]{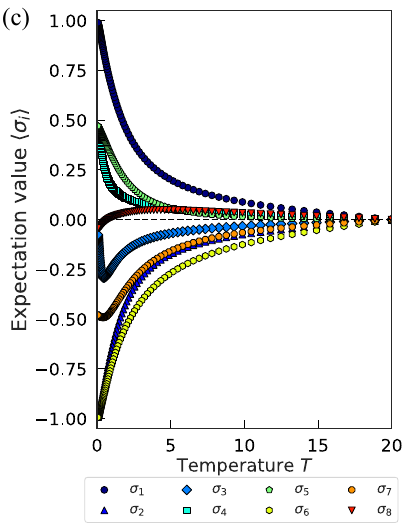}
  \end{minipage}
  \hspace{1cm}
  \begin{minipage}{0.32\linewidth}
    \includegraphics[width=\linewidth]{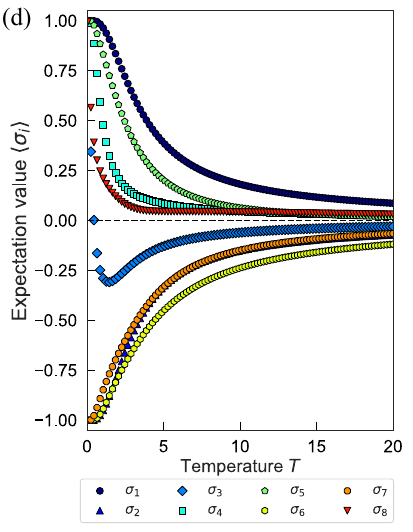}
  \end{minipage}  
  \caption{Dynamic processes and thermal equilibrium states of spin expectation values for Ising model A [(a), (b)] and Ising model B [(c), (d)]. (a), (c) SA performed with $r = 0.555$, (b), (d) Thermal equilibrium states.}
  \label{fig:spin_expect}
\end{figure*}

To investigate the effect of preprocessing SA, we analyzed SA using the master equation from Eq.~\eqref{eq:SA_master_equation}.

We obtain the probabilities of all spin states at time $t$ using the master equation.
Therefore, we calculated the expectation value of spin $\sigma_i$ at time $t$ using the probability of the $k$-th spin state $P_k(t)$, as follows:
\begin{align}
    \langle \sigma_{i} \rangle = \sum_{k=1}^{2^n} P_{k}(t) \sigma_{k, i},
    \label{eq:expectatioin_sigma}
\end{align}
where $\langle \cdot \rangle$ denotes the expectation value, and $\sigma_{k, i}$ denotes $\sigma_i$ in the $k$-th spin state.

This $\langle \sigma_{i} \rangle$ is an indicator for spin fixing when generating the sub-Ising model under the conditions of $N_{\textrm{I}}$ is infinite and $N_{\textrm{S}} = N_{\textrm{I}}$, and it plays the same role as $d_i$ defined in \eqref{eq:d_i}.
When the value of $\langle \sigma_{i} \rangle$ is close to $+1$ or $-1$, the values are more likely to be fixed, and zero indicates that the spins are unstable.

SA was performed using the fourth-order Runge--Kutta method.
In the SA parameters, the initial states were set such that all spin states were equally probable.
The initial temperature $T_{\text{initial}}$ was set to $20$.
The temperature schedule was set to $T(u) = T_{\text{initial}} \times r^u$.
The cooling rate $r$ was set to $0.555$, which was designed under the assumptions of an outer loop of $10$ and a final temperature of $0.1$.
In addition, the equilibrium states at temperature $T$ were analyzed using an exhaustive search.

The spin expectation values for SA and thermal equilibrium states are shown in Fig.~\ref{fig:spin_expect}.
The expectation values of each spin were different when SA was performed (Figs.~\ref{fig:spin_expect}(a) and (c)).
The dynamics of the spin expectation values showed a mixture of spins, where some had expectation values close to $\pm 1$ (stable spins), while others had small values around zero, typically within the range of $-0.25$ to $+0.25$ (unstable spins).
Focusing on spins whose expectation values decreased by the end of SA, these expectation values exhibited low values until low temperatures or nonmonotonic dynamics in the thermal equilibrium analysis~\cite{nakano1968ordering, syozi1968decorated, fradkin1976ising, miyashita1983spin, kitatani1985reentrant, tanaka2005dynamical, miyashita2007nonmonotonic, tanaka2010nonmonotonic} (Figs.~\ref{fig:spin_expect}(b) and (d)). 
This is attributed to the rapid temperature decrease, which leads to trapping in local minima.

These results suggest that stable and unstable spins can be separated, and sub-Ising models are generated with unstable spins by preprocessing SA.
\subsection{Minimum energy gaps of sub-Ising models by QA}
\label{subsec:equation_QA}
To study the characteristics of sub-Ising models when performing QA, the minimum energy gaps of the sub-Ising model were analyzed.
Here, the energy gap refers to the energy difference between the ground state and the first excited state, and the minimum energy gap refers to the smallest energy gap during the annealing process.

\begin{figure}[ht]
  \centering
  \begin{minipage}{0.9\linewidth}
    \includegraphics[width=\linewidth]{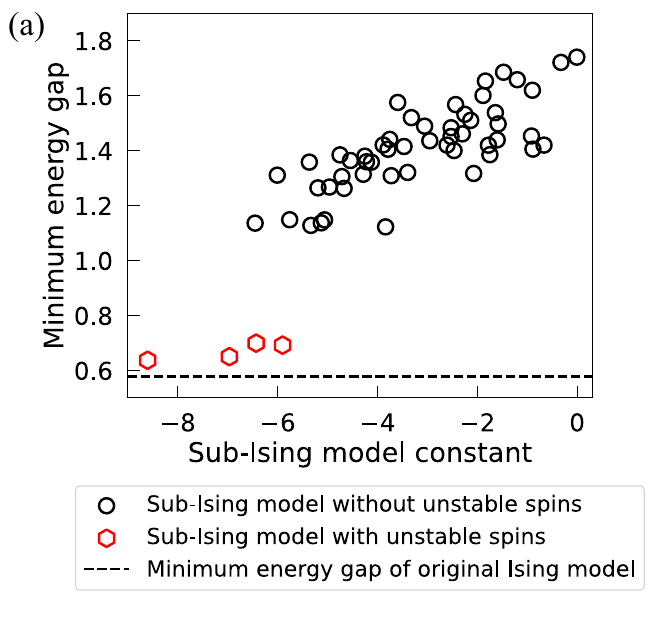}
  \end{minipage}
  \\
  \vspace{5mm}
  \begin{minipage}{0.9\linewidth}
    \includegraphics[width=\linewidth]{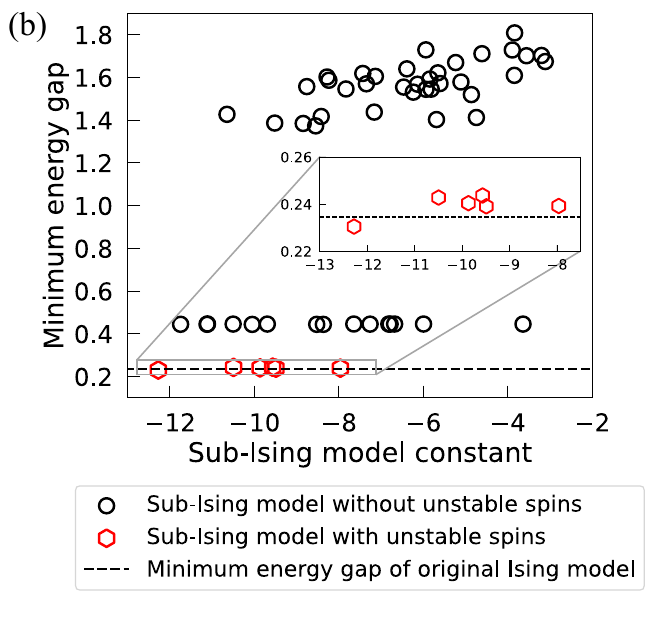}
  \end{minipage}
  \caption{Minimum energy gap and constants of sub-Ising models. (a) The original Ising model is Ising model A, and the sub-Ising model size is five. The unstable spins are $\sigma_1$, $\sigma_2$, $\sigma_5$, and $\sigma_6$ in Fig.~\ref{fig:spin_expect}(a). (b) The original Ising model is Ising model B, and the sub-Ising model size is three. The unstable spins are $\sigma_3$ and $\sigma_8$ in Fig.~\ref{fig:spin_expect}(b). The inset shows a magnified region near the minimum gap of the original Ising model. The black circles, and red hexagons denote the sub-Ising model without and with unstable spins, respectively. Dotted line denotes the minimum energy gap in the original Ising model.}
  \label{fig:energy_gap}
\end{figure}

To evaluate the minimum energy gaps, sub-Ising models were generated according to $\binom{n}{m}$, where $n$ and $m$ denote the original Ising model size and sub-Ising model size, respectively.
For example, when $n = 8$ and $m = 5$, we used a combination of $56$ sub-Ising models.
The fixed spins took the values corresponding to the ground state.

Fig.~\ref{fig:energy_gap} shows the minimum energy gaps and constants of sub-Ising models.
The sub-Ising model constants are described in Eqs.~\eqref{eq:H_sub} and~\eqref{eq:C}.
The black circles denote the results of sub-Ising models without unstable spins and the red hexagons denote those of sub-Ising models with unstable spins.
The unstable spins are identified in Fig.~\ref{fig:spin_expect}.
Most of the minimum energy gaps of sub-Ising models are larger than those of the original Ising model.
It has been reported that fixing spins in a random fully connected Ising model widens the minimum energy gap~\cite{lee2024statistical, hattori2025advantages}.
In addition, the average minimum energy gaps of sub-Ising models with unstable spins widened when the sub-Ising model size was small (Appendix~\ref{sec:appendixA}).
Meanwhile, a new finding revealed that the minimum energy gap of sub-Ising models with unstable spins was smaller than that of sub-Ising models without unstable spins (Fig.~\ref{fig:energy_gap}).
It was also observed that the constant of sub-Ising models with unstable spins is large.
Similar effects were observed for different sizes of the original Ising model.
The results are described in Appendix~\ref{sec:appendixB}.

\begin{figure}[h]
  \centering
  \includegraphics[clip,width=0.8\linewidth]{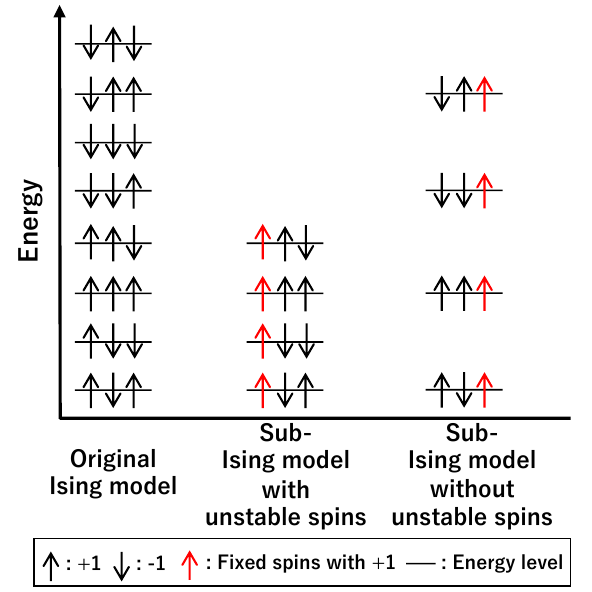}
  \caption{Example of energy spectra with the original Ising model ($n = 3$) and sub-Ising model ($m = 2$). Arrows denote the direction of spins, and the red arrows denote the fixed spins with $+1$.}
  \label{fig:spin_spectra}
\end{figure}

The small minimum energy gap and large constant in sub-Ising models with unstable spins can be attributed to the sub-Ising model generation method.
Fig.~\ref{fig:spin_spectra} shows an example of the energy spectra for the original Ising model and a sub-Ising model. 
The original Ising model size $n$ and sub-Ising model size $m$ are three and two, respectively.
In the sub-Ising model generation method based on sample persistence, spins are fixed when their directions are stable in low-energy solutions.
Therefore, sub-Ising models with unstable spins have only low-energy spectra. 
Because the regions outside the low-energy spectra in sub-Ising models with unstable spins are constant, they have large constants. 
In addition, because the minimum energy gap is within the range of low-energy spectra, the minimum energy gap is smaller.
In contrast, if stable spins are not fixed, the range of the energy spectra becomes larger.
\subsection{Sub-Ising model size dependency}
\label{subsec:size_dependency}
\begin{figure}[ht]
  \centering
  \begin{minipage}{0.85\linewidth}
    \includegraphics[width=\linewidth]{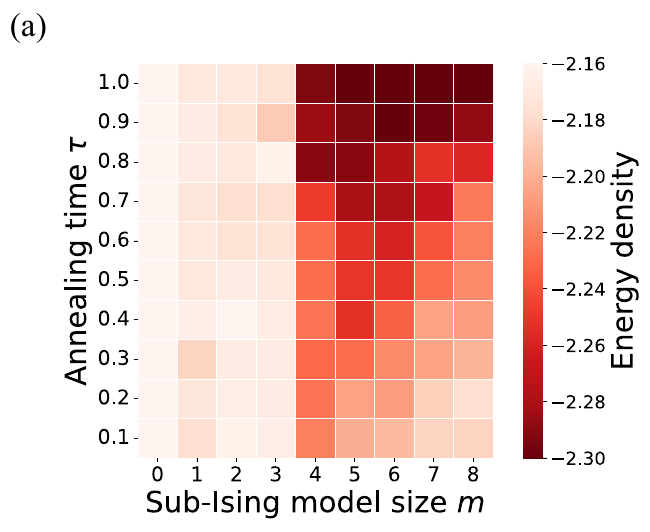}
  \end{minipage}
  \\
  \vspace{5mm}
  \begin{minipage}{0.85\linewidth}
    \includegraphics[width=\linewidth]{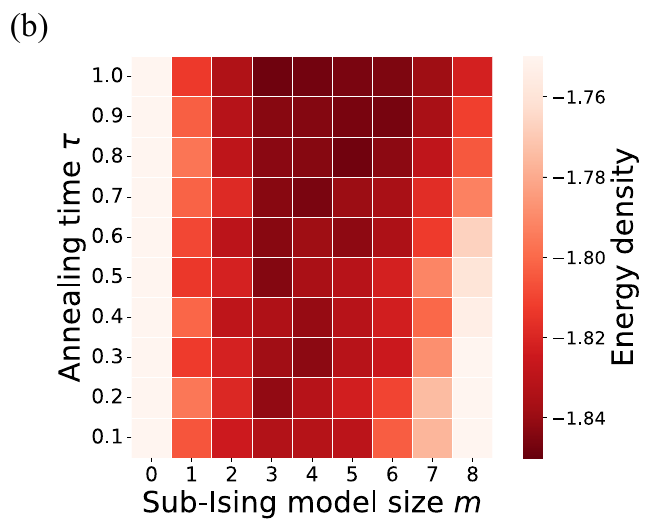}
  \end{minipage}
  \caption{Relationship between the sub-Ising model size and annealing time. The energy densities are the averages of $100$ solutions. (a) Ising model A, and (b) Ising model B. The sub-Ising model size is zero ($m = 0$) shows the results obtained from preprocessing SA.
  For a sub-Ising model size of eight ($m = 8$), the original Ising models were calculated by QA as sub-Ising models.}
  \label{fig:hybrid_SAQA}
\end{figure}

In this subsection, we investigate the sub-Ising model size dependency.
For the analysis, the hybrid optimization method was applied to the small-scale original Ising models used in this section.

Preprocessing SA was implemented using MCMC.
The initial temperature $T_{\text{initial}}$ was set to the value of $\lceil 2v_{\text{max}} \rceil$ from the original Ising model.
The temperature schedule was also similar to that in the previous section, given by \(T(u) = T_{\text{initial}} \times r^u\).
The outer loop was set to five.
QA was performed by evolving the Hamiltonian according to the Schrödinger equation (Subsection~\ref{subsec:QA}) using \verb+mesolve+ in the open-source software \verb+QuTiP+~\cite{johansson2012qutip, johansson2013qutip}. 
QA was used as a quantum annealing machine in the hybrid method.
After performing QA, the probability of each spin state of the sub-Ising model was obtained.
Thus, one spin state is selected based on probability.
To adjust the accuracy of QA, $\tau$ was varied through values $0.1$, $0.2$, $0.3$, $0.4$, $0.5$, $0.6$, $0.7$, $0.8$, $0.9$, and $1.0$ from Eq.~\eqref{eq:H_QA}.
A larger $\tau$ indicates that quantum fluctuations decrease more gradually, thereby increasing the accuracy of QA solutions.

The parameters for the hybrid optimization method were set as $N_{\textrm I} = 10$, $N_{\textrm S} = 5$, $N_{\textrm E} = 10$, and $N_{\textrm L} = 3$. 
SA was executed multiple times to obtain different solution pools, and $100$ sets of solution pools that did not include the ground state were prepared.
The sub-Ising model sizes ranged from one to eight.
The ground state was calculated using an exhaustive search of the spin states.

Fig.~\ref{fig:hybrid_SAQA} shows the average energy density obtained using the hybrid optimization method. 
The part where the sub-Ising model size is zero shows the results obtained from preprocessing SA.
For a sub-Ising model size of eight, the original Ising models were calculated by QA as sub-Ising models.
Therefore, when the sub-Ising model size was eight, the energy densities decreased with increasing annealing time.
When the sub-Ising model size is $1$--$7$, obtaining energy densities lower than those obtained with a sub-Ising model size of eight indicates the effectiveness of the hybrid method.

As shown in Fig.~\ref{fig:hybrid_SAQA}(a), when the sub-Ising model size was three, which is less than the number of unstable spins in Ising model A, the hybrid method was unable to significantly update the solutions obtained by preprocessing SA.
In Fig.~\ref{fig:hybrid_SAQA}(b), the solutions can be updated even when the sub-Ising model size is smaller than the number of unstable spins in Ising model B, which is two.
However, the energy densities were lower when the sub-Ising model size was two or three than when it was one.
These results were not dependent on the annealing time.
This suggests that the sub-Ising model size dependency occurs when the sub-Ising model size is small in relation to the number of unstable spins.
If the fixed spins are different from the ground state, solving the sub-Ising model will not reach the ground state of the original Ising model.
Thus, to reliably obtain the ground state through the hybrid optimization method, it is necessary to set the sub-Ising model size such that all spins with low and unstable expectation values are included in the sub-Ising model. 

Next, we focus on the results for larger sub-Ising model sizes. 
When the annealing time is $0.1$ or $0.2$, the sub-Ising model size shows the lowest energy density at four, as shown in Fig.~\ref{fig:hybrid_SAQA}(a), and three in Fig.~\ref{fig:hybrid_SAQA}(b).
As the sub-Ising model size becomes larger than these values, the energy densities increase. 
In addition, it was observed that the sub-Ising model size with the lowest energy density increased as the annealing time increased.
For example, the sub-Ising model size with the lowest energy density was six, as shown in Figs.~\ref{fig:hybrid_SAQA}(a) and (b), when the annealing time was $0.9$.
These results indicate that the accuracy of the quantum annealing machine affects the size dependency.
In the Ising models used in this study, the average minimum energy gap of sub-Ising models with unstable spins widens as the sub-Ising model size becomes small (Appendix~\ref{sec:appendixA}).
Therefore, as the sub-Ising model size becomes small, the accuracy of QA increases.
In addition, it has been reported that D-Wave machine also shows improved accuracy with smaller problem sizes for fully connected Ising models~\cite{hamerly2019experimental}.
By adjusting the annealing time of D-Wave Advantage, we changed the accuracy of the quantum annealing machine and compared the sub-Ising model sizes to obtain better solutions (see Appendix~\ref{sec:appendixC}).
It was found that the sub-Ising model sizes that obtained better solutions, became larger when the annealing time increased (Fig.~\ref{fig:scatter_appendix}).
This was consistent with the results shown in Fig.~\ref{fig:hybrid_SAQA} and supports our hypothesis.

Therefore, it is suggested that the number of unstable spins and the accuracy of the quantum annealing machine are factors contributing to the sub-Ising model size dependency.
When the sub-Ising model size is small, QA can obtain highly accurate solutions for the model.
However, the fixed spins include those that are not fixed to the values of lower energy, which prevents updating the solutions. 
Meanwhile, when the sub-Ising model size is large, the model includes more spins with low spin expectation values, but it is difficult for QA to obtain high-accuracy solutions.
Therefore, it is important to set a sub-Ising model size to balance this trade-off.

\section{Conclusion and future work}
\label{sec:conclusion}

This study evaluated the performance of a hybrid optimization method for large-scale Ising models that cannot be input into a quantum annealing machine.
Even with large-scale Ising models, the hybrid method updates the solutions from those of preprocessing SA.
In addition, it was observed that the solution updates depended on the sub-Ising model size.

We analyzed the effectiveness and dependency of the sub-Ising model size of the method using numerical simulations of SA and QA. 
In preprocessing SA, it was observed that stable and unstable spins could be separated.
Stable spins were fixed, and unstable spins were included in the sub-Ising model.
Sub-Ising models with unstable spins have lower minimum energy gaps.
However, the minimum energy gaps were still higher than those of the original Ising model.
Moreover, the sub-Ising model dependency is affected by two factors: the number of unstable spins and the accuracy of quantum annealing machines.

This study suggests that the solution accuracy of the quantum annealing machine is crucial for improving the performance of the hybrid optimization method.
It is predicted that as quantum annealing machines can handle larger problem sizes with increased accuracy, the hybrid optimization method will evolve into a high-precision solution approach. 
Improvements in the hardware technology of quantum annealing machines are anticipated.

In this study, we analyzed the characteristics of sub-Ising models using a spin-fixed sub-Ising model generation method based on sample persistence and preprocessing based on SA. 
We speculate that such an analysis is also effective for other preprocessing and sub-Ising model generation methods. 
It is anticipated that the characteristics of sub-Ising model will be similar to those using other preprocessing and sub-Ising model generation methods. 
However, if the characteristics of fixed spins differ for each method, they are expected to contribute to the development of sub-Ising model generation methods.

\appendices
\section{Relationship between minimum energy gaps of sub-Ising models with unstable spins and sub-Ising model size}
\label{sec:appendixA}

In this appendix, we analyze the average minimum energy gaps of sub-Ising models with unstable spins for each sub-Ising model size.
We used the original Ising model described in Section~\ref{sec:analysis_HM}.

\begin{figure}[ht]
  \centering
  \begin{minipage}{0.6\linewidth}
    \includegraphics[width=\linewidth]{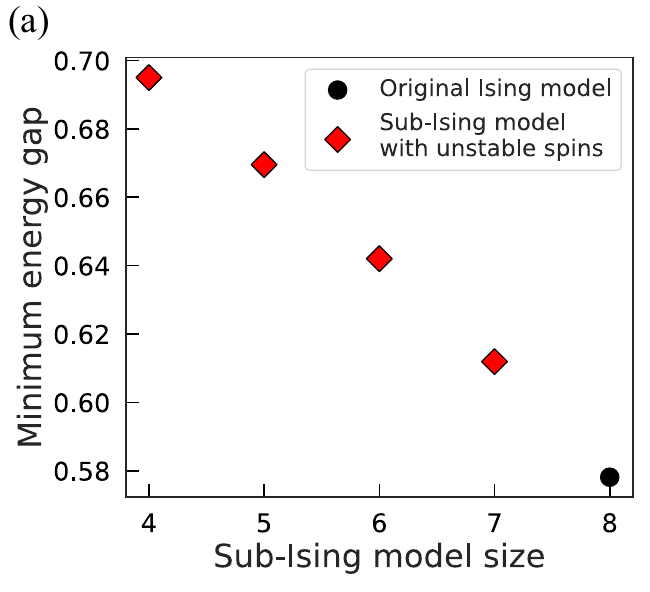}
  \end{minipage}
  \\
  \vspace{5mm}
  \begin{minipage}{0.6\linewidth}
    \includegraphics[width=\linewidth]{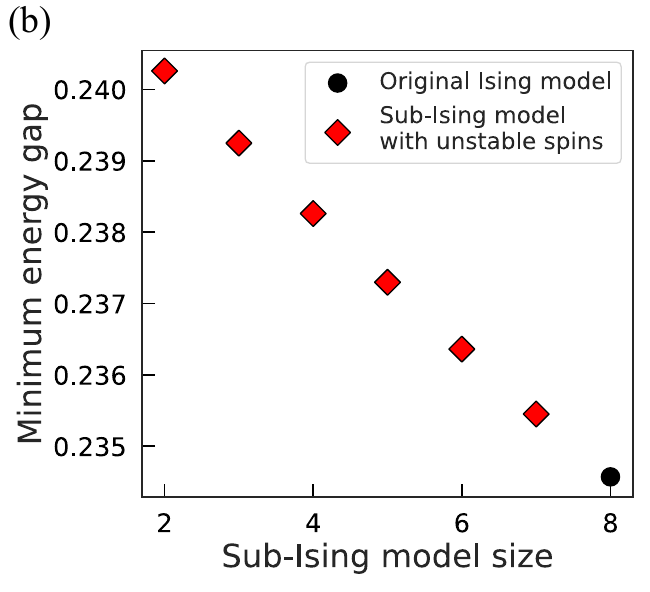}
  \end{minipage}
  \caption{Average minimum energy gap of the sub-Ising models with unstable spins. (a) The original Ising model is Ising model A. Unstable spins are $\sigma_1$, $\sigma_2$, $\sigma_5$, and $\sigma_6$ from Fig.~\ref{fig:spin_expect}(a). The numbers of sub-Ising models are $1$, $6$, $15$, $20$, $15$, and $6$ when the sub-Ising model sizes are $2$, $3$, $4$, $5$, $6$, and $7$. (b) The original Ising model is Ising model B. The unstable spins are $\sigma_3$ and $\sigma_8$ from Fig. ~\ref{fig:spin_expect}(b). The number of sub-Ising models is $1$, $4$, $6$, and $4$ when the sub-Ising model sizes are $4$, $5$, $6$, and $7$. The black circles, and red diamond denote the original Ising model and sub-Ising model with unstable spins. Error bars are not shown in the figure.}
  \label{fig:all_energy_gap_appendix}
\end{figure}

The results are presented in Fig.~\ref{fig:all_energy_gap_appendix}.
The average minimum energy gaps of sub-Ising model with unstable spins widened as the sub-Ising model size decreased.

\section{Original Ising model size differences}
\label{sec:appendixB}

\begin{figure*}[ht]
  \centering
  \begin{minipage}{0.32\linewidth}
    \includegraphics[width=\linewidth]{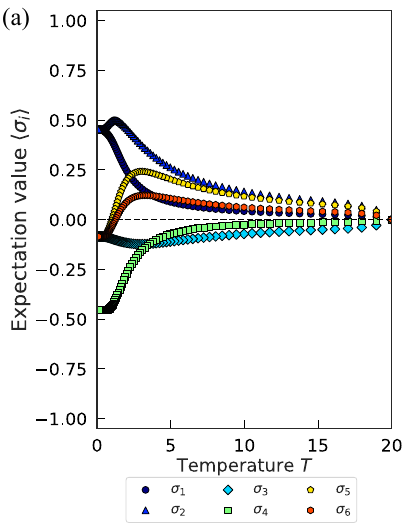}
  \end{minipage}
  \hspace{2.5cm}
  \begin{minipage}{0.32\linewidth}
    \includegraphics[width=\linewidth]{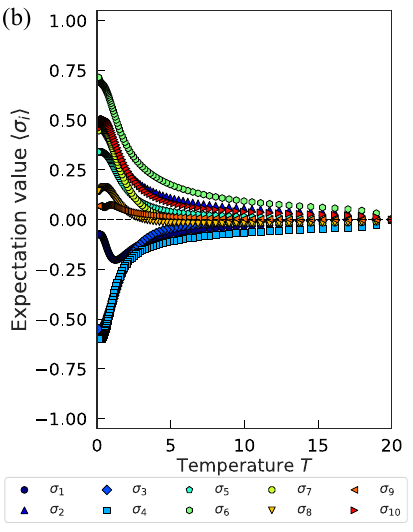}
  \end{minipage}
  \caption{Dynamic process of the spin expectation values when SA was performed with $r=0.948$. Original Ising model sizes with (a) six spins and (b) $10$ spins.}
  \label{fig:spin_expect_appendix}
\end{figure*}

\begin{figure*}[ht]
  \centering
  \begin{minipage}{0.4\linewidth}
    \includegraphics[width=\linewidth]{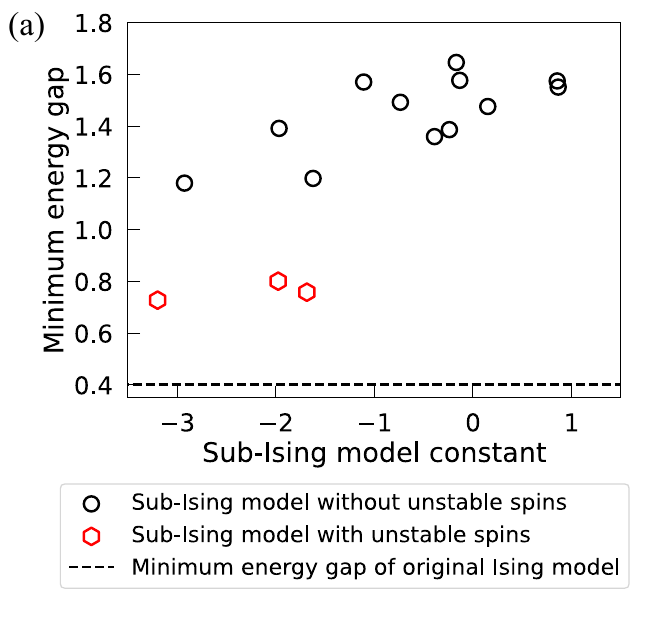}
  \end{minipage}
  \hspace{1cm}
  \begin{minipage}{0.4\linewidth}
    \includegraphics[width=\linewidth]{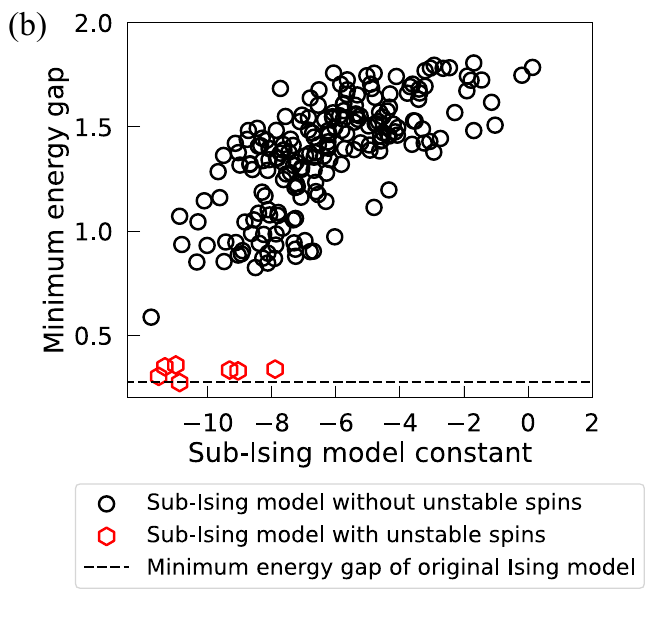}
  \end{minipage}
  \caption{Minimum energy gaps and constants of the sub-Ising models. (a) Original Ising model with $6$ spins. The unstable spins are $\sigma_3$, $\sigma_5$, and $\sigma_6$ in Fig.~\ref{fig:spin_expect_appendix}(a). (b) Original Ising model with $10$ spins. The unstable spins are $\sigma_1$, $\sigma_8$, and $\sigma_9$ in Fig.~\ref{fig:spin_expect_appendix}(b). There were four sub-Ising model sizes. The black circles and red hexagons denote sub-Ising model without or with unstable spins. The dotted line denotes the minimum energy gap in the original Ising model.}
  \label{fig:energy_gap_appendix}
\end{figure*}

In this appendix, the performance of the hybrid method for different Ising model sizes was investigated.
For this purpose, we prepared original Ising models with sizes of $6$ or $10$.
These Ising models are fully connected, and the coefficients were determined by following a Gaussian distribution with a mean of zero and a standard deviation of unity.

First, we analyzed the expectation values for each spin.
For the SA parameters, the initial states, initial temperature, and temperature schedule were set the same as those in Subsection~\ref{subsec:equation_SA}.
The cooling rate $r$ was set to $0.948$, which was designed under the assumptions of an outer loop of $100$ and a final temperature of $0.1$.

The spin expectation values when performing SA are shown in Fig.~\ref{fig:spin_expect_appendix}.
Similar to the results described in Subsection~\ref{subsec:equation_SA}, unstable and stable spins were separated by SA.

Next, we analyzed the minimum energy gaps of sub-Ising models.
The sub-Ising models were generated according to $\binom{n}{m}$, as described in Subsection~\ref{subsec:equation_QA}.
Fig.~\ref{fig:energy_gap_appendix} shows the minimum energy gaps and constants of sub-Ising models.
Similar to Subsection~\ref{subsec:equation_QA}, it was found that sub-Ising models with unstable spins had smaller minimum energy gaps and larger constants than those of sub-Ising models without unstable spins.

\section{Evaluation of the effect of annealing time}
\label{sec:appendixC}

\begin{figure}[ht]
  \centering
  \begin{minipage}{0.9\linewidth}
    \includegraphics[width=\linewidth]{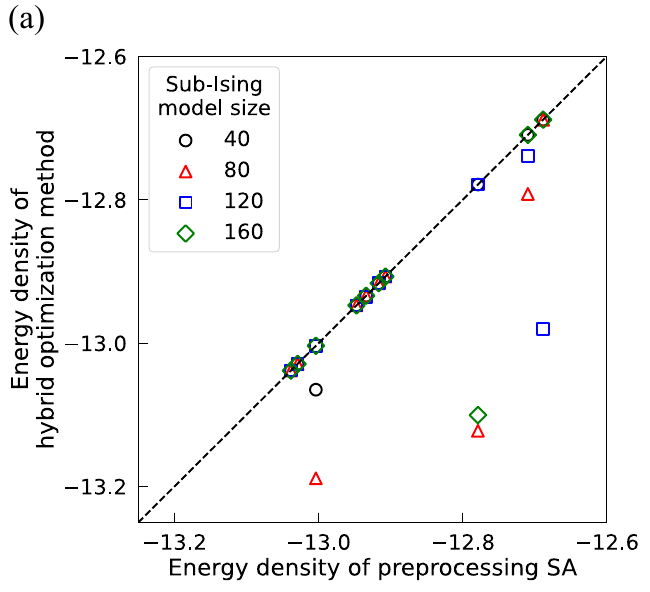}
  \end{minipage}
  \\
  \vspace{5mm}
  \begin{minipage}{0.9\linewidth}
    \includegraphics[width=\linewidth]{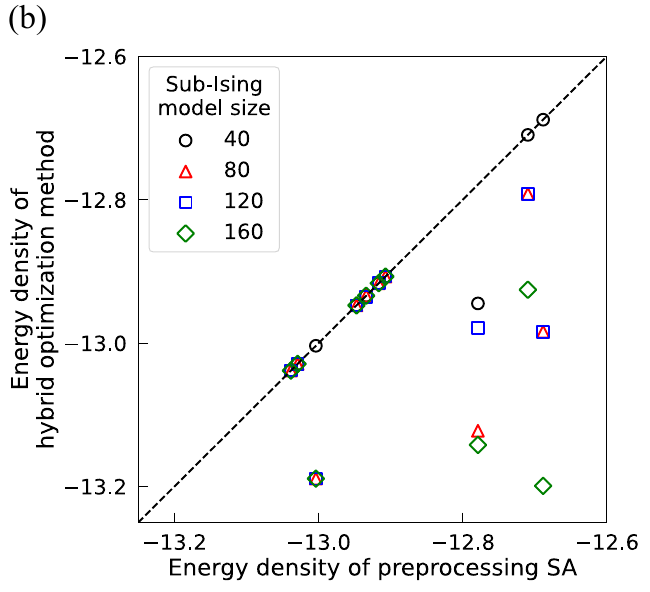}
  \end{minipage}
  \caption{Energy densities of the hybrid optimization method and preprocessing SA, when the annealing time setting changes. The size of the original Ising model was $320$. The annealing times were set to (a) $2$~$\mu$s and (b) $200$~$\mu$s. The black circles, red triangles, blue squares, and green diamonds denote sub-Ising model sizes $m = 40$, $80$, $120$, and $160$, respectively.}
  \label{fig:scatter_appendix}
\end{figure}

In this appendix, we evaluate the effect of the D-Wave Advantage accuracy on the sub-Ising model size dependency.
It is known that the annealing time is a parameter that adjusts the accuracy of D-Wave machine.
Increasing the annealing time improves the accuracy of D-Wave Advantage~\cite{king2023quantum}.
We set the annealing time to $2$~$\mu$s and $200$~$\mu$s.

We prepared an original Ising model with $320$ spins and a solution pool from that used in Section~\ref{sec:evaluation_HM}.
Parameters other than the annealing time for D-Wave Advantage were set to default values~\cite{Ocean}.
For the hybrid method parameters, we set $N_{\textrm I} = 20$, $N_{\textrm S} = 10$, $N_{\textrm E} = 20$, and $N_{\textrm L} = 3$.
The sub-Ising model size $m$ was set to $40$, $80$, $120$, and $160$.

Fig.~\ref{fig:scatter_appendix} shows the energy densities of preprocessing SA and the hybrid method.
When the annealing time was $2$~$\mu$s, the hybrid method tended to update solutions and obtain better solutions with a sub-Ising model size of $80$.
By contrast, when the annealing time was $200$~$\mu$s, solution updates and better solutions were observed with sub-Ising model sizes of $120$ and $160$.

A previous study investigated parameter dependency in the spin reduction method by fixing spins using D-Wave machines~\cite{hattori2025advantages}.
In that study, it was found that when spins were fixed to values different from the ground state, there was an optimal sub-Ising model size.
In addition, it was reported that as the annealing time increased, the optimal sub-Ising model size also increased.
The results in Fig.~\ref{fig:scatter_appendix} are consistent with the findings of this previous study.

\section*{Acknowledgments}
S. Tanaka wishes to express their gratitude to the World Premier International Research Center Initiative (WPI), MEXT, Japan, for their support of the Human Biology-Microbiome-Quantum Research Center (Bio2Q).
The computations in this work were partially performed using the facilities of the Supercomputer Center, the Institute for Solid State Physics, The University of Tokyo.

\bibliography{ref.bib}

\begin{IEEEbiography}[{\includegraphics[width=1in,height=1.25in,clip,keepaspectratio]{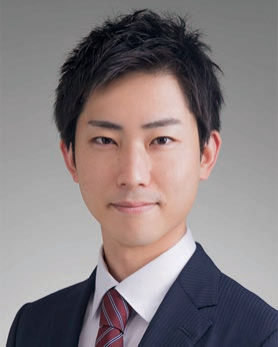}}]{Shuta Kikuchi} received the B.~Eng. and M.~Eng. degrees from Waseda University, Tokyo, Japan, in 2017 and 2019, and Dr.~Eng. degrees from Keio University, Kanagawa, Japan in 2024. He is currently a Project Assistant Professor with the Graduate School of Science and Technology, Keio University. His research interests include Ising machine, statistical mechanics, and quantum annealing. He is a member of the JPS.
\end{IEEEbiography}

\begin{IEEEbiography}[{\includegraphics[width=1in,height=1.25in,clip,keepaspectratio]{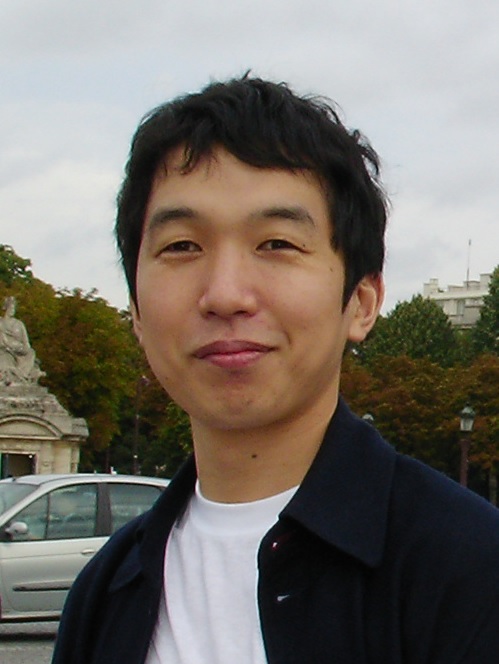}}]{Nozomu Togawa} (Member, IEEE) received the B.~Eng., M.~Eng., and Dr.~Eng. degrees in electrical engineering from Waseda University, Tokyo, Japan, in 1992, 1994, and 1997, respectively. He is currently a Professor with the Department of Computer Science and Communications Engineering, Waseda University. His research interests include quantum computation and integrated system design. He is a member of the ACM, IEICE, and IPSJ.
\end{IEEEbiography}

\begin{IEEEbiography}[{\includegraphics[width=1in,height=1.25in,clip,keepaspectratio]{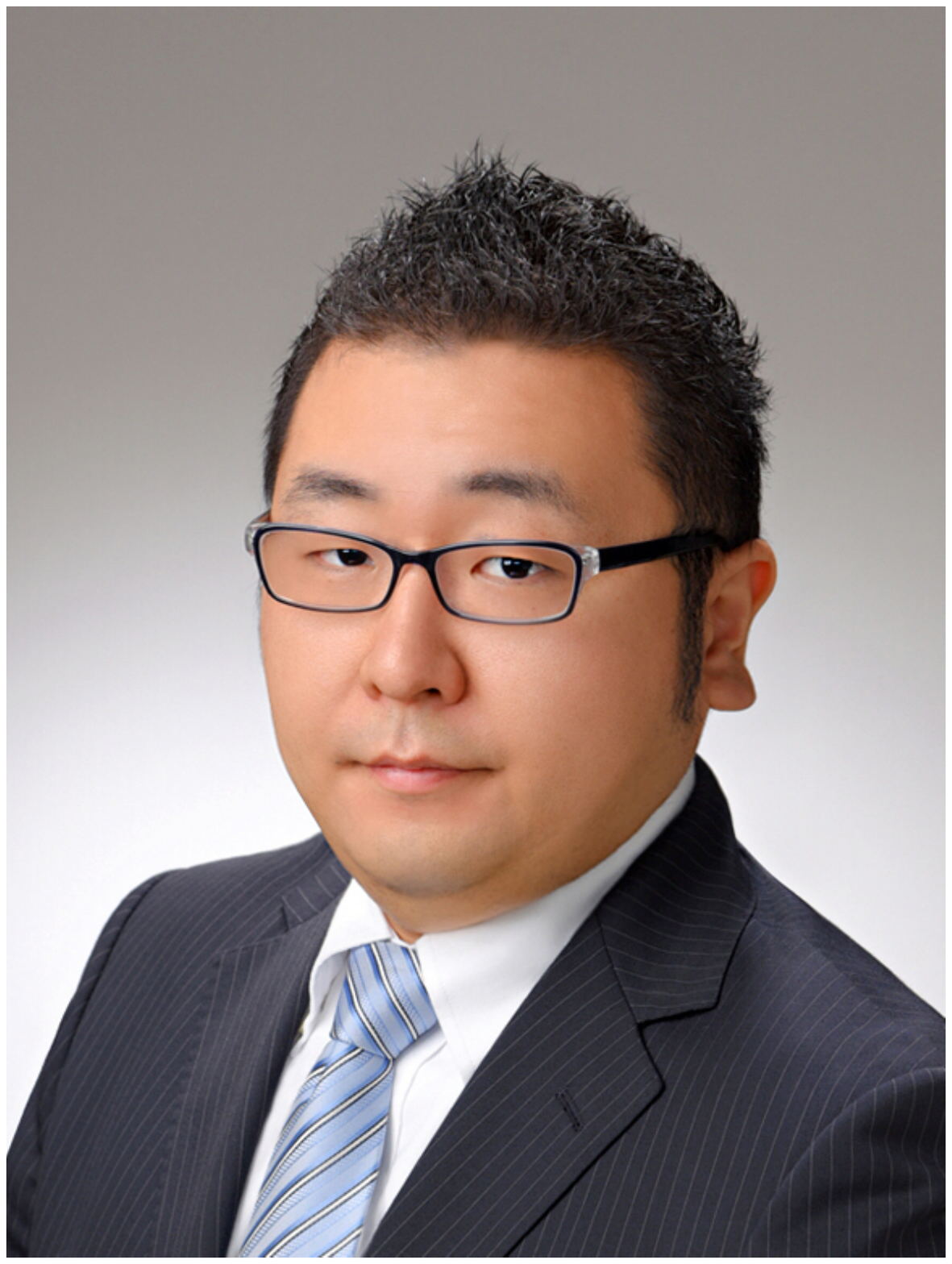}}]{Shu Tanaka} (Member, IEEE) received the B.~Sci. degree from the Tokyo Institute of Technology, Tokyo, Japan, in 2003, and the M.~Sci. and Dr.~Sci. degrees from the University of Tokyo, Tokyo, Japan, in 2005 and 2008, respectively. He is currently a Professor in the Department of Applied Physics and Physico-Informatics, Keio University, a chair of the Keio University Sustainable Quantum Artificial Intelligence Center (KSQAIC), Keio University, and a Core Director at the Human Biology-Microbiome-Quantum Research Center (Bio2Q), Keio University. His research interests include quantum annealing, Ising machines, quantum computing, statistical mechanics, and materials science. He is a member of the Physical Society of Japan (JPS), and the Information Processing Society of Japan (IPSJ).
\end{IEEEbiography}

\EOD

\end{document}